\documentclass{JHEP3}
\newcommand{\be}{\begin{equation}}
\newcommand{\ee}{\end{equation}}
\newcommand{\bea}{\begin{eqnarray}}
\newcommand{\eea}{\end{eqnarray}}
\newcommand{\IR}{\mathbb{R}} 
\newcommand{\IN}{\mathbb{N}}
\newcommand{\IZ}{\mathbb{Z}}
\newcommand{\IS}{{\bf S}}

\newcommand{\non}{\nonumber \\}

\def\half{{1 \over 2}} 

\def\del{{\partial}}
\def\room{~\rule[-2mm]{0mm}{8mm}}

\def\presub{\vspace{.5cm} \noindent}

 \def\tz{\tilde{z}}
 \def\cM{{\cal M}}

\def\tmu{\widetilde{\mu}}

  \def\eps{\epsilon}

\def\room{~\rule[-2mm]{0mm}{8mm}}

\def\Schw{Schwarzschild}
\def\Lp{L_{\rm poles}}
\def\({\left(} \def\){\right)}

\skip\footins = 1\bigskipamount plus 2pt minus 4pt
\usepackage{epsfig}
\usepackage{graphicx}
\def\half{\frac{1}{2}}
\def\del{{\partial}}
\def\room{~\rule[-2mm]{0mm}{8mm}}

\def\Schw{Schwarzschild}
\def\Lp{L_{\rm poles}}
\def\({\left(}
\def\){\right)}

\def\bi{\begin{itemize}}
\def\ei{\end{itemize}}

\preprint{{\tt hep-th/0505009}}
\title{Matched Asymptotic Expansion for Caged Black Holes -
Regularization of the Post-Newtonian Order}

\author{Dan Gorbonos and  Barak Kol\\
Racah Institute of Physics, Hebrew University\\
Jerusalem 91904, Israel\\
E-mail: \email{gdan@phys.huji.ac.il},
{\tt\href{mailto:barak_kol@phys.huji.ac.il}{barak\_kol@phys.huji.ac.il}}}

 \abstract{The ``dialogue of multipoles'' matched asymptotic
expansion for small black holes in the presence of compact
dimensions is extended to the Post-Newtonian order for arbitrary
dimensions. Divergences are identified and are regularized through
the matching constants, a method valid to all orders and known as
Hadamard's partie finie. It is closely related to ``subtraction of
self-interaction''
 and
shows similarities with the regularization of quantum field
theories. The black hole's mass and tension (and the ``black hole
Archimedes effect'') are obtained explicitly at this order, and a
Newtonian derivation for the leading term in the tension is
demonstrated. Implications for the phase diagram are analyzed,
finding agreement with numerical results and extrapolation shows
hints for Sorkin's critical dimension -- a dimension where the
transition turns second order.}

\begin{document}
 \section{Introduction}

The black-string black-hole transition (which includes the
Gregory-Laflamme instability \cite{GL1}) is a phase transition in
General Relativity which occurs at higher dimensions $d>4$ where
black hole uniqueness fails. It is known to raise deep issues
including topology change and critical dimensions \cite{TopChange}
as well as naked singularities and thunderbolts (see the reviews
\cite{review,HO-review} and references therein).

In order to test these issues, black object solutions were sought
both analytically and numerically in $\IR^{d-2,1} \times \IS^1$.
One limit which is amenable to analytic study is the small black
hole limit, where $\rho_0$, the black hole radius, is much smaller
than $L$, the size of the compact dimension. Admittedly, small
black holes are far from being the large black holes ($\rho_0 \sim
L$) which are involved in the phase transition and the latter
should be studied numerically. Yet, they are still very useful in
providing the following \bi
 \item Hints about large black holes via extrapolation.
 \item Important tests for numerics.
 \ei

In \cite{previous} we introduced a general perturbation theory for
small black holes, an implementation of the method of ``matched
asymptotic expansion'' to the static case, which was termed ``a
dialogue of multipoles'' since it describes how field multipoles
change the black hole's shape (or mass multipoles) and these in
turn change the field and so on. This method requires to obtain a
solution in two zones: the near horizon zone and the asymptotic
zone, solutions which must agree over the overlap region.

In principle it should be possible to devise a perturbation theory
in a single zone. For instance, Harmark wrote an approximate
solution in a single zone \cite{H4}, but nobody extended it to a
full perturbation series. However, it is probably simpler to use
the two-zone method since it does not involve the arbitrariness of
an initial guess, and since the differential operator to be
inverted is the same at each order.

In \cite{previous} we obtained the leading order corrections to
the metric. Recently a ``matched asymptotic expansion'' was used
to obtain the explicit solutions in 5d up to the next to leading
order ( ${\cal O}(\rho_0^{~4}),\,{\cal O}(L^{-4})$ ) \cite{KSSW1}
(see also \cite{KSSW0} for a closely related work in the
braneworld context).

In this paper we extend our method to the next to leading order in
the asymptotic zone, namely to Post-Newtonian order, for arbitrary
$d$ (namely, order $2(d-3)$). It turns out that at this order a
new qualitative phenomenon appears -- \emph{divergences}. Such
divergences are familiar from Post-Newtonian studies, but we are
not aware of a general prescription that works at all orders. In
section \ref{divergences} we obtain such a clear regularization in
terms of a ``cut-off and match'' method, which relies on
Hadamard's ``Partie finie'' regularization and is summarized in
equation (\ref{regular-multi}). This method shows some similarity
with the regularization of quantum field theories (see
\cite{Goldberger-Rothstein} for related ideas). Alternatively, a
certain subtraction of self-interaction is shown to be an
equivalent method at this order. We believe this equivalence
should continue to hold at arbitrary orders, but this is not
manifest in the current setting. Moreover, we find that at this
order the matching is quite trivial and matching constants vanish.
Again we do not know whether this can be made to persist to higher
order and whether it depends on a clever choice of gauge order by
order.

The above-mentioned divergences did not show up in neither
\cite{H4} nor \cite{KSSW1}: \cite{H4} was using a different,
single patch, method while in \cite{KSSW1} the authors succeeded
in obtaining full explicit solutions for the metric without the
use of Green's function, thereby circumventing the issue of
divergences. Still, the regularization of these divergences is
important both conceptually and practically: conceptually, they
are a general feature of the two-zone method and show up at all
high enough orders; and practically, they are essential if one
wishes to use Green's function either while resorting to a
numerical solutions for the whole metric, or when computing
analytically the solutions' asymptotics, namely the
thermodynamics.
 Moreover, similar divergences are very common
in Post-Newtonian studies in general.

The pre-existing results for the leading corrections are reviewed
in section \ref{rev}. Proceeding to next to leading order, we met
difficulties in trying to obtain explicit expressions for the full
metric, but we were able to compute the next to leading
corrections to the black hole's mass and tension, using our
regularization procedure, thereby confirming the results of
\cite{H4} with our different method. The computation is described
in section \ref{PN-thermo} and the results are presented in
subsection \ref{results}. The leading behavior of the tension is
given an intuitive Newtonian explanation in subsection
\ref{Newtonian-tension}. A higher order expression for the ``black
hole Archimedes effect'' is given in Appendix \ref{Archimedes}.

Finally, we discuss the implications for the phase diagram  in
section \ref{implications}, in light of the objectives presented
in the beginning of this introduction. Within the small black hole
region of validity we find very good agreement with numerical data
in 5d and 6d \cite{KPS1,KudohWiseman1,KudohWiseman2}.
Extrapolating beyond this region we find evidence for a second
order phase transition at large enough dimensions, which is an
indication for Sorkin's critical dimension \cite{SorkinD*}.

\subsection{Set-up}

\begin{figure}[htb] \epsfxsize=0.6\textwidth

\centerline{\epsffile{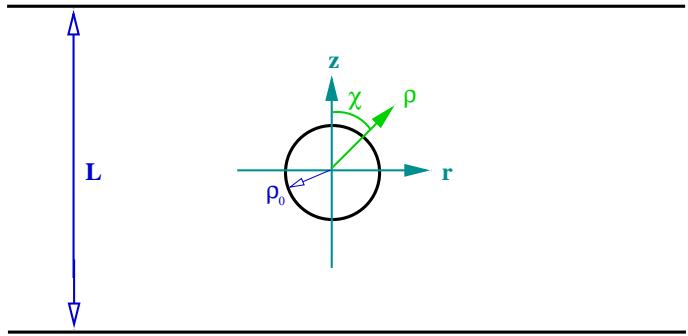}} \caption{\small
Illustration of the $(r,z)$ ``cylindrical" coordinates and the
$(\rho,\chi)$ ``spherical" near horizon coordinates. The period of
the compact dimension (in the $z$ direction) is denoted by $L$ and
the radius of the black hole is denoted by $\rho_{0}$.}
 \label{PiCcoordinates}
\end{figure}

Figure \ref{PiCcoordinates} illustrates the coordinates which we
use. We denote by $z,\, r$ the ``cylindrical'' coordinates, where
$z$ is the coordinate along the compact dimension whose period we
denote by $L$, and $r$ is the radial coordinate in the extended
$\IR^{d-2}$ spatial dimensions. In addition we introduce
``spherical'' coordinates $\rho,\, \chi$. $\rho_0$ is a
characteristic size of the horizon. We are interested in obtaining
solutions for small black holes, $\rho_{0} \ll L$, so $\rho_{0}/L$
is the small parameter. The ``spherical'' coordinates will be more
natural in the vicinity of the black hole since it is nearly
spherical in the small black-hole limit.

\begin{figure}[htb]
\epsfxsize=0.7\textwidth \centerline{\epsffile{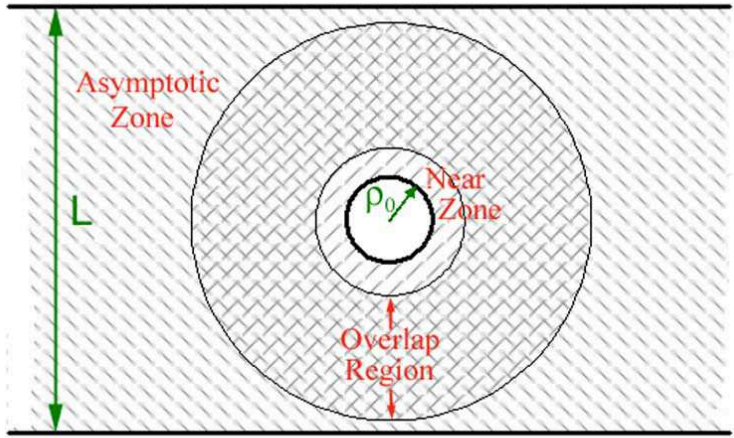}}
\caption{\small The division of the spacetime into two zones: the
\textbf{near zone} $\rho \ll L$ where $\rho_0$ is fixed and the
perturbative parameter is $L^{-1}$, and the \textbf{asymptotic
zone} $\rho \gg \rho_{0}$ where $L$ is fixed and the perturbative
parameter is $\rho_{0}$. The two zones overlap over the
\textbf{overlap region} which increases indefinitely in the small
black hole limit. During the perturbation process the two zones
are separate, and communicate only through the matching
``dialogue''. The near zone is defined by $\{ (\rho,\chi): ~ \rho
\ge \rho_0 \}$ while the asymptotic zone is defined by $\{ (r,z):~
r \ge 0, ~z \sim z + L \} \backslash (0,0)$.} \label{zones}
\end{figure}

The general method of ``dialogue of multipoles'' matched
asymptotic expansion was explained in \cite{previous}. The static
nature of the problem under study makes this application of
matched asymptotic expansion more transparent than the usual 4d
applications. In this method we consider two zones (see figure
\ref{zones}): \bi
\item The \emph{asymptotic zone}
where $\rho \gg \rho_{0}$ and $\rho_{0}$ is the small parameter.
The zeroth order solution is flat space with a periodic coordinate
$z \sim z+ L$ and the point at the origin $(r,z)=(0,0)$ removed.
 \item The \emph{near zone} where $\rho \ll L$ and $L^{-1}$ is the
small parameter. The zeroth order solution is the \Schw ~black
hole with radius $\rho_0$. \ei
 Thus in the asymptotic zone we
expand the metric in $\rho_{0}$ and in the near zone we expand it
in $L^{-1}$ \bea
 g_{\mu\nu}^{(\mbox{asymp})}(x) = \sum_{j=0}^{\infty}\,
 \rho_0^{~j}\,
 g_{\mu\nu}^{(\mbox{asymp},j)}(x)~, \non
 g_{\mu\nu}^{(\mbox{near})}(x) = \sum_{j=0}^{\infty}\, L^{-j}\,
 g_{\mu\nu}^{(\mbox{near},j)}(x) ~.\eea
Orders which are integral multiples of $d-3$ have a special role,
and therefore we introduce a special square brackets notation for
them \be
 g_{\mu\nu}^{[k]}:=g_{\mu\nu}^{\(k(d-3) \)} ~.\ee

The metric in the two regions must be consistent over the overlap
region $\rho_0 \ll \rho \ll L$, in a double expansion in
$\rho_0,\, L$. This is the ``matching procedure'', whose first
steps are summarized in figure \ref{ddialogue}. For example, the
first match is between the \Schw-Tangherlini solution in the near
zone (order $L^{-0}$) and the Newtonian approximation (order
$\rho_{0}^{d-3}$).

 \section{Divergences and regularization}
\label{divergences} \noindent {\bf Divergences}. When we proceed
in the ``dialogue of multipoles'' matched asymptotic expansion
beyond the leading order we encounter a new qualitative feature --
divergences. These occur in both the asymptotic zone and the near
zone, and for concreteness we shall first describe the divergences
and their regularization for the asymptotic zone in detail, and
later the near zone will be described briefly.

At each order, $k$, in the perturbation series we get an equation
of the form \be
 L\, h^{(k)} = Src^{(k)}
 ~, \label{pert-eq} \ee
 where $L$ is the linear operator which appears at first order,
$h^{(k)}$ is a function which determines the perturbation to the
metric at order $k$ and $Src^{(k)}$ is a source term which depends
on metric functions from lower orders. More specifically, in the
asymptotic zone we have \be
 \triangle\, h^{(k)} = Src^{(k)} \sim \del^{k_{1}}\, \Phi^{k_{2}} ~, \label{pert-eq-asymp} \ee
 where $\triangle$ is the Laplacian, $\Phi$ is the Newtonian
potential given explicitly by \be \Phi :=
\rho_{0}^{d-3}\sum_{n=-\infty}^{\infty} \frac{1}
{\left(r^{2}+(z+nL)^{2}\right)^{\frac{d-3}{2}}}, \label{thephi2}
~, \ee
 $\del^{k_{1}}$ is a symbolic notation for some differential
operator of order $k_1$ and $k_{1},k_{2}\ge 2$. \footnote{This
form allows for non-linearities arising from both the Newtonian
approximation $\sim \Phi$ and sources introduced through multipole
matching $\sim \del^l \Phi$. $k_{2} \ge 2$ since the source is
non-linear and $k_{1} \ge 2$ from a dimensional analysis.}
 For small $\rho$ $\Phi$ behaves as  $\Phi \sim 1/\rho^{d-3}$ and
for large $r$ as $\Phi \sim 1/r^{d-4}$. For example the
Post-Newtonian equations (PN) are \bea
 \triangle \( h_{tt}^{[2]} + \half\, \Phi^2 \) &=& 0 ~,\non
 \triangle \( h_{ij}^{[2]} - \half\, \delta_{ij}\, \({\Phi \over d-3}\)^2 \) &=&
  - {d-2 \over 2(d-3)}\, \Phi_{,i}\, \Phi_{,j} \label{PN} ~,\eea
where $i,j$ run over the spatial indices.

Usually we solve an equation such as (\ref{pert-eq-asymp}) by
means of a Green's function. Namely, we take the solution to be
\be
 h_G(x) := \int dx'\, G(x,x')\, Src(x') ~, \label{Green-eq} \ee
 where in this case \footnote{We take the Green function with no boundary conditions at
 0. One could have taken a more involved $G(x,x')$ with an arbitrary
 multipole distribution at $x=0$, but this would not cure the divergences which
 we will encounter. For the precise prefactor in (\ref{G-Phi}) see (\ref{Green-func}).}
\be
 G(x,x') \propto \Phi(x-x') ~.\label{G-Phi} \ee

Divergences of the integral may come from regions where the
integrand is infinite or from large regions of integration. In our
case short-distance divergences are possible in the vicinity of
$x' \to 0$ (there are none as $x' \to x$), and long-distance
divergences are possible as $r \to \infty$. By a standard abuse of
language we shall refer to the short-distance divergences as ``UV
divergences'' and to long-distance ones as IR even though the
theory is not quantum and distances and energies are not
reciprocal.

UV divergences. The integrand of (\ref{Green-eq}) diverges as
$\rho' \to 0$ due to the singularity in $Src(\rho')$ which is
inherited from $\Phi(\rho')$. Assuming that for small $\rho'$ the
behavior of the singular and angle-independent (``S-wave'') part
of the integrand is $G(x,x')\, Src(x') \sim \rho'^{-s}$ we get
from (\ref{Green-eq}) \be
 h_G(x) \simeq \int_0 d\rho' \rho'^{d-2} \,
\rho'^{-s} ~. \ee
 Therefore we have UV divergences (for all $x$) exactly if $s \ge d-3$.
 For PN we see from (\ref{PN}) that $Src(\rho') \simeq
\Phi_{,i}\, \Phi_{,j} \simeq \rho^{-2(d-2)}$, \footnote{More
precisely, one needs to find the degree of the singular and
angle-independent part of $G(x,x')\, Src(x')$. Since for PN the
leading singularity of $Src(\rho')$ has a quadrupole part then $s
\ge 2(d-3)$ and the conclusion is unchanged.} and therefore
\emph{we indeed have UV divergences} for all $d$. At higher orders
the divergences only get worse.

IR divergences. If for large $r$ $Src$ behaves as $Src(r) \sim
r^{-s_\infty}$ we get \be
 h_G(x) \simeq \int^\infty dr' r'^{d-3} \, {1 \over r'^{d-4}} \,
r'^{-s_\infty} ~. \ee Therefore we would have divergences exactly
if $s_\infty \le 2$, but the slowest relevant decay rate of $Src$
-- a PN source in $d=5$ -- is $s_\infty=2(d-3)=4$ and hence
\emph{there are no IR divergences}.

\presub {\bf Unregulated multipoles}. Even though the Green
function integral diverges we can still find a particular solution
of (\ref{pert-eq-asymp}). Locally around the origin we may
separate the equation by using radial variables $(\rho,\chi)$ and
we find \be
 \( \del_\rho^{~2} + {d-2 \over \rho}\, \del_\rho -{l(l+d-3) \over
 \rho^2} \) \, h_l=Src_l ~.\ee
The particular solution is \be
 h_l \propto \rho^{-s_l+2} \label{particular} ~,\ee
 where the constants $s_l$ are defined through $Src_l(\rho) \sim \rho^{-s_l}$, and
 one should allow for a possible addition of a homogeneous solution of the form
$\rho^l,~ \rho^{-(d-3)-l}$.

Indeed, if particular solutions can be found in explicit form, and
their homogeneous part determined by matching, then there is no
problem of divergences, as was done in \cite{KSSW1} for $d=5$.
However, we are often interested only in the asymptotic form of
the metric and then a Green-function-based multipole expansion is
useful. For instance, in this paper we will compute only the
corrections to the thermodynamic quantities, the mass and the
tension, rather than the whole metric. Rephrasing from a different
perspective, the particular solution (\ref{particular}) at small
$\rho$ does not give information about the large $\rho$
asymptotics, and the separation of variables is not valid
globally.

We now wish to re-formulate the divergence problem in terms of
multipoles, which are particularly suited for divergences
localized at a point. Since the divergences come from the vicinity
of $x'=0$ we Taylor expand $G(x,x')$ there. The convergence radius
$\rho_{cnvg}$ would be $\sim \min(L,x)$ and for $x \to \infty$
$\rho_{cnvg} \sim L$. Then we separate the domain of integration
in (\ref{Green-eq}) to within the convergence radius
 $\rho<\rho_{cnvg}$ and outside it. Outside there are no
divergences, and inside we may perform a standard multipole
expansion \bea
 h(x) \simeq \del'_{i_1} \dots \del'_{i_l} G(x,0)\,
 \int_{\rho<\rho_{cnvg}} dV'\, Src(x')\, x'^{i_1} \dots x'^{i_l}
 ~.\eea
Thus we see that all the divergences may be encoded by the
diverging multipoles at the origin, defined by the small $\eps$
behavior of \be
 {\cal M}_l:= \int_{|\rho|>\eps} dV\, Src(x)\, x^{i_1} \dots x^{i_l}~. \label{div-multi} \ee

\presub {\bf Regularization}. The matching boundary conditions
(b.c.) suggest a natural regularization procedure: ``cut-off and
match''. Namely, cut-off the integral (\ref{div-multi}) at
$\rho=\eps$, adjust it by a constant (corresponding to adding an
allowed homogeneous piece to $h$) to conform with the matching
b.c., and finally take the limit $\eps \to 0$.

This idea is translated into formulae as follows. Expand the
multipoles in a Laurent series in $\eps$ \be
 \cM_l(\eps)=\sum_{j=j_{min}}^{\infty}\, \cM_{l,j}\, \eps^j  ~, \label{multi-Laurent} \ee
where at order $k$ dimensional analysis gives
$j_{min}=-(k-l-(d-3))$.
 From the finite part multipoles, $\cM_{l,0}$, we may
construct a function $h$ which is known in the mathematical
literature as Hadamard's partie finie \cite{Hadamard}, or
Hadamard's finite part, and hence we shall denote \be
 \mbox{Pf}(\cM_l):=\cM_{l,0} \ee
 (see also \cite{BlanchetFaye-Hadamard} for its use in the Post-Newtonian
context). Alternatively, the regularization could be defined
through analytic continuation and it is also closely related to
Cauchy's Principal Value.

The divergent piece in (\ref{multi-Laurent}), (namely the sum over
$j<0$), is interpreted as coming from the vicinity of the origin,
and thus according to the matching b.c. should be replaced by an
appropriate value of the multipole $\cM^{match}_l$ read from the
asymptotics ($\rho \gg \rho_0$) of the near zone metric. Thus we
arrive at the expression for \emph{the regularized multipoles}
which is one of our central formulae
\be
\fbox{$~\room \cM_l \to \cM_l^{~R} =\mbox{Pf}(\cM_l) +
\cM^{match}_l$ }
 ~.\label{regular-multi} \ee

\presub {\bf ``Renormalization''}. This regularization procedure
is reminiscent of renormalization in Quantum Field Theory. There
we know that in order to renormalize a theory it must have as many
free parameters, or counter-terms, to be set by experimental data,
as divergences.\footnote{In QFT a theory is called normalizable if
the number of counter-terms is finite, and non-renormalizable
otherwise. In our case there will be infinitely many divergences,
but all the necessary ``experimental data'' will come from
matching.} Here by analogy we need \be
 \#(\mbox{divergences}) = \#(\mbox{matchings}) ~. \label{renorm-condition} \ee

However, at each order there are only certain (finitely many)
matchings coming from the other zone, as encoded by the
``perturbation ladder'' (see figure \ref{ddialogue}). What would
guarantee that (\ref{renorm-condition}) is satisfied? The answer
turns out to be reminiscent of Quantum Field Theory as well -- it
is dimensional analysis! At order $k$ in the asymptotic zone we
are considering divergences of the form \be
  {\rho_0^{~k} \over \rho^{l+d-3}\, \eps^a\, L^b} ~,\ee
 where $a>0,\, b \ge 0$. Since all the quantities in this expression
have dimensions of length, but the expression itself (namely $h$)
is dimensionless then $a+b=k-l-(d-3)$ and in particular
$k-(d-3)-l>0$. However, this is exactly the condition for
matching, as is seen by considering appropriate terms at order $b$
in the near zone, namely taking $a=0,\, b>0$.

\presub {\bf No self-interaction}. We are also familiar with a
different approach to regularization -- the removal of
self-interaction. The idea is to separate the black hole solution
into a Schwarzschild piece and a correction coming from the
compact dimensions. Intuitively one expects that all the
divergences come from the singular \Schw ~part, and that these
will be absorbed automatically by adjusting its parameters.

This approach is commonly used in the Post-Newtonian context (see
for example the review talk \cite{Poisson-GR17}). However, it is
not known how to implement this idea at arbitrary orders of
perturbation, but rather some results are known at low orders. We
expect the ``no-self-interaction'' to be equivalent to our method
and it would be interesting to demonstrate it. Our explicit
calculations in the next chapters do lend support to this idea, as
it turns out that if one omits singular products from the double
sum over images in PN, it is equivalent to Hadamard's
regularization. In other words, Hadamard's partie finie of the
singular part is zero at this order.

Moreover, we observe another simplification whereby the matching
is exactly cancelled by the multipole contribution of the ${\cal
O}(\Phi^2)$ term within the Laplacian in (\ref{PN}). This may very
well be a special feature of our gauge, and it would be
interesting to determine whether it can be made to continue in
higher orders as well. So altogether in our PN computation
regularization may be replaced by discarding the singular
self-interaction source term and matching is automatically
cancelled.

\presub {\bf The near zone}. In the near zone the regularization
of divergences works in parallel to the method just described for
the asymptotic zone, so we only describe the main points.
Actually, in this paper we shall not treat the near zone at all.

Here the linear operator, $L$, in (\ref{pert-eq}) is given by the
Heun equation (see ~\cite{previous}), which for $\rho \gg \rho_0$
reduces once more to the Laplacian. The divergences are IR rather
than UV. Regularization proceeds by introducing a large-distance
cut-off $R$ and considering the Laurent series for multipoles as
$R \to \infty$. After a partie finie regularization the matched
value from the asymptotic zone should be added just as in
(\ref{regular-multi}).

\section{Review of leading order results}
\label{rev}
\subsection{The general procedure}

\begin{figure}[htb]
\epsfxsize=0.7\textwidth \epsfysize=0.45\textwidth
 \centerline{\epsffile{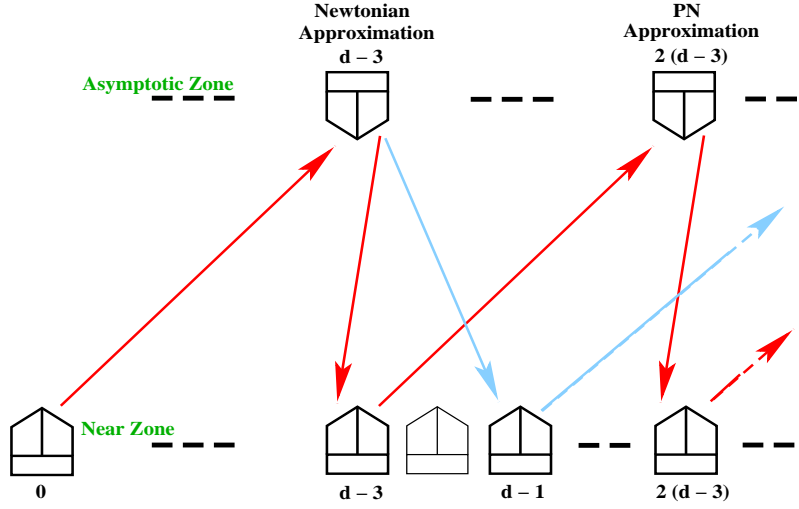}}
\caption{\small The first steps in the matching procedure. Each
box in the top row denotes a specific order in the asymptotic
zone, and the lower row depicts the near zone. Arrows denote the
flow of matching information between the zones: Dark (red) arrows
denote monopole matching ($l=0$),  light arrows (light-blue)
denote the quadrupole ($l=2$), and higher $l$ arrows are not
shown.}
 \label{ddialogue}
\end{figure}

Using the matching procedure described in \cite{previous} we
obtained the leading corrections to the \Schw ~metric of a black
hole as a result of the compact dimension. The matching procedure
for the leading corrections is summarized in figure
\ref{ddialogue}.

The metric in the asymptotic zone is expanded in the radius of the
black hole $\rho_{0}$ in the following form
\begin{equation}
g_{\mu\nu}^{\mbox{(asymp)}}=\eta_{\mu\nu}+h_{\mu\nu}^{[1]}+h_{\mu\nu}^{[2]}+...,
\end{equation}
where $h_{\mu\nu}^{[1]} \propto \rho_{0}^{d-3}$,$h_{\mu\nu}^{[2]}
\propto \rho_{0}^{2\,(d-3)}$ and so on. Note that the matching
procedure may introduce corrections which are not integer powers
of $\rho_{0}^{d-3}$ but this would happen only after the
post-newtonian corrections.

According to the matching procedure, in order to determine the
Post-Newtonian corrections (order $2\,(d-3)$ in figure
\ref{ddialogue}) we first need to know the Newtonian approximation
and the near zone leading order. Therefore we start here by
reviewing the results that were obtained in ~\cite{previous} for
these orders. The leading order in the asymptotic zone is simply
the Newtonian potential while the leading order corrections in the
near zone are decomposed into multipolar static perturbations to
the ~\Schw-Tangherlini metric (following Regge-Wheeler
\cite{Regge} and using a similar gauge).

For the post-Newtonian expansion we make a quite-standard gauge
choice. One starts by writing the Ricci tensor as ~\cite{fock2}
\begin{equation}
R_{\mu\nu}=-\frac{1}{2}g_{\mu\sigma}g_{\nu\rho}g^{\alpha\beta}
\frac{\partial^{2}g^{\rho\sigma}}{\partial x^{\alpha}\partial
x^{\beta}}+\Gamma_{\mu}^{\alpha\beta}\Gamma_{\nu,\alpha\beta}-\Gamma_{\mu\nu},
\label{posteq}
\end{equation}
where $\Gamma_{\mu,\alpha\beta}$ and $\Gamma_{\mu}^{\alpha\beta}$
are the Christoffel symbols of the first and the second kind,
respectively, and in addition one defines
\begin{eqnarray}
\Gamma^{\nu} &:=& g^{\alpha\beta}\Gamma^{\nu}_{\alpha\beta}\,,
\nonumber\\
\Gamma_{\mu\nu}&:=&
\frac{1}{2}\bigg(g_{\mu\rho}\frac{\partial\Gamma^{\rho}}{\partial
x^{\nu}}+ g_{\nu\sigma}\frac{\partial\Gamma^{\sigma}}{\partial
x^{\mu}}- g_{\mu\rho}g_{\nu\sigma}\frac{\partial
g^{\rho\sigma}}{\partial x^{\alpha}}\Gamma^{\alpha} \bigg) \,.
\nonumber
\end{eqnarray}
Next one chooses the Harmonic (or de Donder~\footnote{The first
introduction of this gauge appeared in~\cite{deDon}.}) gauge by
the requirement that
\begin{equation} \label{gaugeE}
\Gamma^{\nu}=\Box
x^{\nu}=\frac{1}{\sqrt{-g}}\frac{\partial}{\partial
x^{\beta}}(\sqrt{-g} g^{\beta\nu}) \equiv 0\, ,
\end{equation}
where we denote by $g$ the determinant of the metric $g_{\mu\nu}$.
In this gauge, the last term in the expression of the Ricci tensor
above vanishes.

\subsection{Newtonian approximation}
\label{Newtonsec} The first step in the iterative procedure was to
solve the equations in the Newtonian approximation (see for
example \cite{previous,H4,HO1})
\begin{eqnarray}
h_{tt}^{[1]}&=&\Phi, \label{Newton1}\\
h_{ij}^{[1]}&=&\frac{1}{d-3}\,\Phi\,\delta_{ij}, \label{Newton2}
\end{eqnarray}
and \be \Phi := \rho_{0}^{d-3}\sum_{n=-\infty}^{\infty} \frac{1}
{\left(r^{2}+(z+nL)^{2}\right)^{\frac{d-3}{2}}}, \label{thephi}\ee
 where the Latin indices stand for the spatial components.
In order to match this approximation with the near zone we have to
expand $\Phi$ around $\rho=0$ in polar coordinates
$(\rho,\chi)$
\be \label{expansion}
\Phi(r,z)=\frac{\rho_{0}^{d-3}}{\rho^{d-3}}+2\cdot
\frac{\rho_{0}^{d-3}}{L^{d-3}}\,\zeta(d-3)+
\frac{(d-3)\,(d-2)\,\rho^{2}\,\rho_{0}^{d-3}}{L^{d-1}}\,\zeta(d-1)\,\Pi^{2,d}_{0}(\chi)+\mathcal{O}(\rho^{4}),
\ee where $\zeta$ is Riemann's zeta function\footnote{Riemann's
zeta function is defined as
$\zeta(s)=\sum_{n=1}^{\infty}\frac{1}{n^{s}}$.} and
\[\Pi^{2,d}_{0}(\chi):=\frac{1}{d-2}\,\left(\cos^{2}(\chi)\,(d-1)-1\right),\]
which is the generalized Legendre function that corresponds to the
quadrupole. In particular, the overall prefactor in (\ref{thephi})
is set by the matching described by the left most arrow in figure
\ref{ddialogue}.

From the asymptotic form of the metric (monopole terms), when $r
\gg z$, \be
g_{\mu\nu}=\frac{c_{\mu\nu}}{r^{d-4}}+\mathcal{O}\(\frac{1}{r^{d-3}}\),
\ee one can extract the mass and tension \cite{KPS1,HO2} defined
through the standard first law \be
 dM={\kappa \over 8 \pi}\, dA + \tau dL ~. \label{first-law} \ee
(For weak sources this definition is equivalent to $M:=\int
d\,V_{d-1}T_{tt},\, \tau:=-\int d\,V_{d-1}T_{zz}/L$.)

In harmonic gauge the asymptotic measurable quantities can be
expressed in terms of the constants $c_{\mu \nu}$ in the
asymptotics as follows \cite{KPS1,HO2} \bea
M=\frac{\Omega_{d-3}}{16\,\pi\, G_{d}}\((d-3)\,c_{tt}-c_{zz}\), \label{mass}\\
\tau=\frac{\Omega_{d-3}}{16\,\pi\, G_{d}\,
L}\(c_{tt}-(d-3)c_{zz}\), \label{tens} \eea where
\[\Omega_{d-3}=\frac{2\,\pi^{\frac{d-2}{2}}}{\Gamma(\frac{d-2}{2})},\]
is the area of a unit $S^{d-3}$.

At this order, the asymptotic constants can be determined from the
asymptotic expansion\footnote{Which may be gotten from flux
conservation for $\overrightarrow{\nabla}\Phi$.}
 ($r \gg z$) of the Newtonian potential

 $\Phi$
 \be \Phi=\frac{\Omega_{d-2}}{\Omega_{d-3}} \cdot
\frac{d-3}{d-4}\,
\rho_{0}^{d-3}\,\frac{1}{r^{d-4}}+\mathcal{O}\(\frac{1}{r^{d-3}}\).
\label{Phiasym} \ee

Then \bea
c_{tt}^{[1]}=\frac{\Omega_{d-2}}{\Omega_{d-3}} \cdot \frac{d-3}{d-4}\, \rho_{0}^{d-3},  \\
c_{zz}^{[1]}=\frac{\Omega_{d-2}}{\Omega_{d-3}}
\cdot\frac{1}{d-4}\, \rho_{0}^{d-3}. \eea

Thus, the tension vanishes in the first order approximation and
for the mass we obtain

\be M=\frac{(d-2)\Omega_{d-2}}{16
\pi\,G_{d}}\,\rho_{0}^{d-3}+\mathcal{O}\(\rho_{0}^{2(d-3)}\).
\label{firstmass} \ee

\subsection{The near zone metric}
\label{near-zone-rev}

In the near zone we obtained the first monopole correction to the
\Schw ~metric by matching with the Newtonian
approximation~\cite{previous}. The metric in the near zone reads
\be
\begin{array}{ccc}

 \room g_{tt,d}^{\mbox{near}}&=-&\left(1-\frac{\rho_{0}^{d-3}}{\rho_{s}^{d-3}}\right)\left(1-\frac{2\,\zeta(d-3)\,\rho_{0}^{d-3}}{L^{d-3}}\right)+\mathcal{O}\left(\frac{1}{L^{d-2}}\right), \label{final_mono}\\
\room g_{\rho\rho,d}^{\mbox{near}}&=&
\left(1-\frac{\rho_{0}^{d-3}}{\rho_{s}^{d-3}}\right)^{-1}
 \left( 1+\frac{2\,\zeta(d-3)\,\rho_{0}^{d-3}}{(d-3)\,L^{d-3}}
 \right) +\mathcal{O}\left(\frac{1}{L^{d-2}}\right),\\
  \room
  g_{\chi\chi,d}^{\mbox{near}}&=&\rho_{s}^{2}\left(1+\frac{2\,\zeta(d-3)\,\rho_{0}^{d-3}}{(d-3)\,L^{d-3}}\right)+\mathcal{O}\left(\frac{1}{L^{d-2}}\right).
\end{array}  \ee

The metric in the near zone is written in \Schw ~coordinates when
the metric in the asymptotic zone is written in Harmonic
coordinates (ones which satisfy (\ref{gaugeE})). We denote by
$\rho_{s}$ the radial coordinate in \Schw ~coordinates and by
$\rho$ the radial coordinate in the Harmonic coordinates. In order
to match them we have to use the leading terms of the
transformation rule~\cite{previous}
\begin{equation} \label{gauge_ex}
\rho=\rho_{s}-\frac{\rho_{0}^{d-3}}{2\,(d-3)\,\rho_{s}^{d-4}}+\mathcal{O}(\frac{1}{\rho_{s}^{d-3}}).
\end{equation}
and get the near zone metric in Harmonic coordinates \bea
 g_{tt}^{\mbox{near}}&=&-\(1-\frac{\rho_{0}^{d-3}}{\rho^{d-3}}\)
 \(1-\frac{2\,\zeta(d-3)\,\rho_{0}^{d-3}}{L^{d-3}}\)+\mathcal{O}
 \left(\frac{1}{\rho^{2(d-3)}},\frac{1}{L^{d-2}}\right), \label{near1}\\
g_{rr}^{\mbox{near}}&=&g_{zz}^{\mbox{near}}=\(1+\frac{\rho_{0}^{d-3}}
{(d-3)\,\rho^{d-3}}\)\(1+\frac{2\,\zeta(d-3)\,\rho_{0}^{d-3}}{(d-3)\,L^{d-3}}\)
+\mathcal{O}\left(\frac{1}{\rho^{2(d-3)}},\frac{1}{L^{d-2}}\right).\nonumber
\eea

Note that the prefactor of $\frac{1}{\rho^{d-3}}$ in the near zone
determines the monopole corrections to the asymptotic zone metric
through the matching procedure. In particular, the term that
behaves as $\frac{\rho_{0}^{2(d-3)}}{\rho^{d-3}\,L^{d-3}}$ is the
one that should be matched with the post-Newtonian corrections to
the metric in the asymptotic zone (see figure \ref{ddialogue}).

\presub {\bf Higher multipoles}.

From the ``perturbation ladder'' (figure \ref{ddialogue}) we see
that proceeding in the near zone beyond the leading correction
(order $d-3$)  we have contributions of higher multipoles  that
are determined from the Newtonian potential, before the first
non-linear iteration at order $2(d-3)$. The multipole moments of
the newtonian potential are matched with the multipole linear
perturbations of the BH. In our case, only even multipole numbers
contribute corrections to the metric due to the $\chi \rightarrow
\pi -\chi$ symmetry. Any even multipole number $l<d-3$ contributes
a correction to order $l+d-3$ in $L^{-1}$ before order $2\,(d-3)$
(see \cite{previous}), where the next monopole correction enters.
For example, in $d=5$ there are no corrections of higher
multipoles before order 4, in $d=6$ we have a quadrupole
correction $l=2$ and so on.

For any $l<d-3$ we obtained in \cite{previous} explicit
expressions for the correction to the near zone metric. We review
here the final results. The correction of order $l+d-3$ can be
expressed in the form \be
g_{\mu\nu}^{(\mbox{near},l+d-3)}\,dx^{\mu}\,dx^{\nu}=g_{tt}^{\mbox{(near},l+d-3)}\,dt^{2}+g_{\rho_{s}\rho_{s}}^{(\mbox{near},l+d-3)}\,d\rho_{s}^{2}+g_{\chi\chi}^{(\mbox{near},l+d-3)}\,d\Omega_{d-2}^{2},\ee
where \bea
g_{\chi\chi}^{(\mbox{near},l+d-3)}&=&\rho_{s}^{2}\,c_{l}\,\rho_{0}^{l}\frac{\Gamma(1+\frac{l}{d-3})\,\Gamma(2+\frac{l}{d-3})}{\Gamma(1+\frac{2\,l}{d-3})\,(l-1)}\,E_{l}(X)\,\Pi_{0}^{l,d}(\chi),\\
E_{l}(X)&=&\(1-(d-3)\,X\,\frac{d}{d\,X}\)\((1-X)\,_{2}\!F_{1}(1-\frac{l}{d-3},2+\frac{l}{d-3},2\,;1-X)\),\nonumber\\
X&:=&\frac{\rho_{s}^{d-3}}{\rho_{0}^{d-3}}, \eea $_{2}\!F_{1}$ is
the hypergeometric function and $\Pi_{0}^{l,d}(\chi)$ are the
generalized Legendre polynomials given by a Rodriguez formula
\begin{equation} \label{rodrig}
\Pi^{l,d}_{0}(\chi)=\frac{\Gamma(\frac{d}{2}-1)}{\Gamma(l+\frac{d}{2}-1)}\,\sin^{4-d}(\chi)\,\left(\frac{1}{\sin(\chi)}\,\frac{d}{d\chi}\right)^{l}\,\sin^{2l+d-4}(\chi).
\end{equation}
The constants $c_{l}$ come from the metric in the asymptotic zone,
namely, from the Newtonian potential. These are the constants that
appear in  \be
\frac{g_{\chi\chi}^{(\mbox{asymp},d-3)}}{\rho^{2}}=\frac{\Phi}{d-3}=
\frac{1}{d-3}\,\frac{\rho_{0}^{d-3}}{\rho^{d-3}}+
\sum_{l=0}^{\infty}\!c_{l}\,\rho^{l}\,\Pi_{0}^{l,d}(\chi)~, \ee -
see (\ref{expansion}) for the explicit values of $c_l$ for
$l=0,2$.
 The other two components of the metric correction are
determined from $g_{\chi\chi}^{(\mbox{near},l+d-3)}$ \be
\begin{array}{ccc}
g_{tt}^{(\mbox{near},l+d-3)}&=&c_{l}\,\rho_{0}^{l}\,\frac{\Gamma(1+\frac{l}{d-3})\,\Gamma(2+\frac{l}{d-3})}{\Gamma(1+\frac{2\,l}{d-3})\,(l-1)}\,f\,A_{l}(\rho_{s})\,\Pi_{0}^{l,d}(\chi),
\\ \\
g_{\rho\rho}^{(\mbox{near},l+d-3)}&=&c_{l}\,\rho_{0}^{l}\,\frac{\Gamma(1+\frac{l}{d-3})\,\Gamma(2+\frac{l}{d-3})}{\Gamma(1+\frac{2\,l}{d-3})\,(l-1)}\,\frac{B_{l}(\rho_{s})}{f}\,\Pi_{0}^{l,d}(\chi),
\end{array} \quad \quad l\,<\,d-3
 \ee
where \bea f &:=& 1-\frac{\rho_{0}^{d-3}}{\rho_{s}^{d-3}},\\
A_{l}(\rho_{s})&=&\frac{(d-2)\,\rho_{s}\,f}{2\,l\,(l+d-3)}\left[(d-2)\,\frac{d\,E_{l}}{d\rho_{s}}+\rho_{s}\,\frac{d^{2}\,E_{l}}{d\rho_{s}^{2}}\right]+\frac{d-4}{2}E_{l},
\nonumber \eea and $B_{l}$ is obtained from the algebraic relation
\[B_{l}=-A_{l}-(d-4)\,E_{l}.
\]
\label{higher}

\section{Post-Newtonian Thermodynamics}
\label{PN-thermo}

\subsection{Post-Newtonian equations}

Using the results for $h_{\mu\nu}^{[1]}$ (the Newtonian
approximation) we can write Einstein's equations for
$h_{\mu\nu}^{[2]}$, i.e. the Post-Newtonian equations, in the
following form \bea
\triangle \(h_{tt}^{[2]}+\frac{1}{2}\Phi^{2}\)=0,      \label{post}  \\
\triangle
\(h_{ij}^{[2]}-\frac{1}{2}\delta_{ij}\(\frac{\Phi}{d-3}\)^{2}\)&=&
-\frac{d-2}{2(d-3)}\Phi_{,i}\Phi_{,j} ~~. \nonumber \eea The
origin contains a singular source which will be accounted for by
matching rather than by introducing a point-like source.

Let us look at the ``source'' term (the RHS) for the spatial
components. This term is proportional to
\[\Phi_{,i}\Phi_{,j}=(d-3)^{2}\rho_{0}^{2\,(d-3)}
\sum_{m,n \in \IZ}\frac{\(x_{i}+m\,L\,\delta_{iz}\)\,\(x_{j}+n\,L\,\delta_{jz}\)}
{\(r^{2}+(z+m\,L)^{2}\)^{\frac{d-1}{2}}\,\(r^{2}+(z+n\,L)^{2}\)^{\frac{d-1}{2}}}.\]
In this sum we see the effect of interaction between the black
hole and its mirror images, including the interaction between the
mirror images themselves. One term in this sum is different - the
term that corresponds to the self-interaction of the black hole,
namely, when $m=n=0$. This term exists in the post-Newtonian
equations regardless of the boundary conditions. It has a singular
behavior near the origin being proportional to
$\frac{x_{i}x_{j}}{\rho^{2\,(d-1)}}$ . For $i=j$ its integral over
the volume diverges like $1 \over \varepsilon^{d-3}$ for
$\varepsilon \rightarrow 0$.

We shall follow the ``no self-interaction'' approach together with
``matching regularization'' (see the discussion in section
\ref{divergences}). Since the equations are linear we can separate
the equations to the self-interaction part (SI) and the regular
part (REG) which consists of the interaction with the mirror
images. In order to keep the periodic boundary conditions in the
compact dimension, we include in the SI part the self-interaction
terms of the mirror images even though they are not singular in
the domain $|z| \le L/2$.

Accordingly we separate \be h_{\mu\nu}^{[2]}=h_{\mu\nu}^{[2]\,
SI}+h_{\mu\nu}^{[2]\, REG}. \ee
 The separated PN equations are \bea
\triangle \(h_{tt}^{[2]\,SI}+\frac{1}{2}\Phi^{2}_{SI}\)&=&0,       \label{sing1} \\
\triangle \(h_{ij}^{[2]\, SI
}-\frac{1}{2\,(d-3)^{2}}\delta_{ij}\Phi^{2
}_{SI}\)&=&-\frac{d-2}{2(d-3)}\(\Phi_{,i}\Phi_{,j}\)_{SI},
\label{sing2}
 \eea
where
\[\Phi^{2 }_{SI}=\rho_{0}^{2(d-3)}\,\sum_{n\in \IZ} \frac{1} {\(r^{2}+(z+n\,L)^{2}\)^{d-3}},\]
\[\(\Phi_{,i}\Phi_{,j}\)_{SI}=(d-3)^{2}\rho_{0}^{2\,(d-3)}\sum_{n\in \IZ}\frac{\(x_{i}+n\,L\,\delta_{iz}\)\,\(x_{j}+n\,L\,\delta_{jz}\)}{\(r^{2}+(z+n\,L)^{2}\)^{d-1}},\]
 and \bea
\triangle \(h_{tt}^{[2]\,REG}+\frac{1}{2}\Phi^{2}_{REG}\)&=&0,       \label{reg1} \\
\triangle \(h_{ij}^{[2]\, REG
}-\frac{1}{2\,(d-3)^{2}}\delta_{ij}\Phi^{2
}_{REG}\)&=&-\frac{d-2}{2(d-3)}\(\Phi_{,i}\Phi_{,j}\)_{REG},
\label{reg2}
 \eea
\newpage
where
\[\Phi^{2 }_{REG}=\rho_{0}^{2(d-3)}\,\sum_{m \neq n} \frac{1}
{\(r^{2}+(z+m\,L)^{2}\)^{\frac{d-3}{2}}\,\(r^{2}+(z+n\,L)^{2}\)^{\frac{d-3}{2}}},\]

\[\(\Phi_{,i}\Phi_{,j}\)_{REG}=(d-3)^{2}\rho_{0}^{2\,(d-3)}\sum_{m
\neq
n}\frac{\(x_{i}+m\,L\,\delta_{iz}\)\,\(x_{j}+n\,L\,\delta_{jz}\)}
{\(r^{2}+(z+m\,L)^{2}\)^{\frac{d-1}{2}}\,\(r^{2}+(z+n\,L)^{2}\)^{\frac{d-1}{2}}},\]
\[\lefteqn{\(\, m,n \in \IZ  \,\).} \]
 Finally, the boundary conditions determining the homogeneous
solution are supplied through matching with the near zone metric
(\ref{near1}), as usual.

\presub {\bf The self interaction part -- SI}. An explicit form of
a particular solution for the singular part
(\ref{sing1},\ref{sing2}) can be found. For this purpose let us
look at the equations in the case of a single source without any
mirror images ($n=0$), whose metric we denote by
$h_{\mu\nu}^{[2]\, 0}$ \bea \triangle \(h_{tt}^{[2]\,
0}+\frac{\rho_{0}^{2(d-3)}}{2\, \rho^{2(d-3)}}\)&=&0,
\label{sing10}
\\
\triangle \(h_{ij}^{[2]\, 0}-\frac{1}{2
\,(d-3)^{2}}\frac{\rho_{0}^{2(d-3)}}{
\rho^{2(d-3)}}\delta_{ij}\)&=&-\frac{(d-2)\,(d-3)}{2}\,\rho_{0}^{2(d-3)}\,\frac{x_{i}\,x_{j}}{\rho^{2\,(d-1)}}.
\label{sing20} \eea

The solution for the equations above
 is
 \bea
h_{tt}^{[2]\,0}&=&-\frac{\rho_{0}^{2(d-3)}}{2\, \rho^{2(d-3)}},
\label{sing10sol}
\\
h_{ij}^{[2]\,0}&=&\rho_{0}^{2\,(d-3)}\left[\frac{\delta_{ij}}{2\,(d-3)^{2}\,\rho^{2\,(d-3)}}
+\left\{\!\!\begin{array}{lr} \frac{\ln(\rho)}{8\,
\rho^{4}}\(\frac{4\,x_{i}\,x_{j}}{\rho^{2}}-\delta_{ij}\)
+\frac{x_{i}\,x_{j}}{12\,\rho^{6}}-\frac{11\,\delta_{ij}}{96\,\rho^{4}}
\quad d=5 \non
 \frac{1}{4\, \rho^{2\,(d-3)}\,(d-5)}\(\frac{\delta_{ij}}{d-3}
 -\frac{x_{i}\,x_{j}\,(d-3)}{\rho^{2}}\)
\quad d>5
\end{array}
\right.\right].\eea

Now, the solution to the full SI equations
(\ref{sing1},\ref{sing2}) can be obtained by taking $z \rightarrow
z+n\,L $ in (\ref{sing10sol}) and summing over $n\in \IZ$ \bea
h_{tt}^{[2]\,SI}&=&\sum_{n=-\infty}^{\infty}\!h_{tt}^{[2]\,0}(r,z+n\,L),\\
h_{ij}^{[2]\,SI}&=&\sum_{n=-\infty}^{\infty}\!h_{ij}^{[2]\,0}(r,z+n\,L).
\eea

\presub {\bf The regular part -- REG}. The particular solution in
the asymptotic zone can be expressed by an integral over Green's
function \bea
 h_{tt}^{[2]\,REG}&=&-\frac{1}{2}\Phi^{2}_{REG}, \label{partasym}\\
 h_{ij}^{[2]\, REG}&=& \frac{1}{2\,(d-3)^{2}}
\,\delta_{ij} \Phi^{2 }_{REG}-\frac{d-2}{2\,(d-3)}
  \int \(\Phi_{,i}\Phi_{,j}\)_{REG}(r',z') G(r-r',z-z')
 d\,V_{d-1},
\nonumber
\label{partasym2} \\
 \eea
where Green's function is \be G(r-r',z-z')=-\frac{1}{(d-3)
\Omega_{d-2}}\,\sum_{n=-\infty}^{\infty}
\frac{1}{\((r-r')^{2}+(z-z'+n\,L)^{2}\)^\frac{d-3}{2}}.
\label{Green-func} \ee

\subsection{Matching}

We still need to match the solutions with the near zone, a
matching which will turn out to vanish. As encoded in the
``ladder'' figure (figure \ref{ddialogue}) we should match the
monopole term at order $2(d-3)$ in the asymptotic zone, with the
monopole of order $d-3$ in the near zone. Accordingly, the
relevant term in the double expansion in the overlap region
$\rho_0 \ll \rho \ll L$ is \be
\frac{\rho_{0}^{2(d-3)}}{\rho^{d-3}\,L^{d-3}}, \label{theterm1}
\ee

The matching ``budget'' is composed of several items summarized in
the table \ref{table1}, which we proceed to explain.
\begin{table}[t]
\be \begin{array}{c|cc}
             & h_{tt} & h_{zz} \\
\hline
 \room SI           & 0      & 0 \\
 \room REG ~\Phi^2 \mbox{ shift} & -\half\, c_\Phi & +{1 \over 2\,(d-3)^2}\, c_\Phi  \\
 \room REG \mbox{ Green's} & 0      & 0 \\
 \hline \hline
 \room \mbox{order } [1] \mbox{ near} & -2\, \zeta &+{2\, \zeta \over (d-3)^2}  \\
\end{array}
\ee
 \caption{Matching budget for both $h_{tt}$ and $h_{zz}$ at Post-Newtonian order,
 namely for the coefficients of ${\cal O}\(\rho_{0}^{2(d-3)}/(\rho^{d-3}\,L^{d-3})\)$.
 At PN these quantities have 3 contributions
 SI -- Self-interaction solutions see (4.9); 
 REG $\Phi^2$ shift -- namely the contribution from $\Phi^{2
}_{REG}$ to the regular term see (4.12,4.13,4.17). 
 Here $c_\Phi:=4 \zeta$ is the coefficient of the relevant term in $\Phi^{2
}_{REG}$ (4.17) and in this table $\zeta := \zeta(d-3)$;
 REG Green's - for $h_{zz}$ the Green's function integral in (4.13) 
 has no relevant term according to (4.18). 
 For $h_{tt}$ this term is identically zero (4.12). 
 In the fourth line we see the leading order results for the near zone, taken
from (3.17), 
 and we confirm that they are identical to the sum
 of the 3 items above them and therefore the matching constants
 vanish.}
 \label{table1}
\end{table}

\bi \item SI --- In the case of the Self-interaction equations we
have an explicit solution (\ref{sing10sol}). One sees that neither
$h_{tt}^{[2]}$ nor $h_{zz}^{[2]}$ contain any term of the form
(\ref{theterm1}).

In the regular solutions (\ref{partasym},\ref{partasym2}), we
distinguish two parts, the first term is just a shift proportional
to  $\Phi^2_{REG}$, which we refer to as ``REG $\Phi^2$ shift'',
and the second is an integral with a Green function kernel (there
is no such term in (\ref{partasym}) ), which we refer to as ``REG
Green's''.

\item REG $\Phi^2$ shift ---
 In $\Phi^{2 }_{REG}$ we have a term of the relevant form
(\ref{theterm1})
  \be
  \Phi^{2 }_{REG}=\frac{4 \,
\rho_{0}^{2\,(d-3)}} { \rho^{d-3} \, L^{d-3}} \, \zeta(d-3)+
\frac{4 \, \rho_{0}^{2\,(d-3)}}{L^{2\,(d-3)}} \,\(
\zeta^{2}(d-3)-\frac{1}{2}\,\zeta\(2\cdot(d-3)\)\)+...,
\label{phiex}\ee whose coefficient we denote by $c_\Phi \equiv 4\,
\zeta(d-3)$,
 where we used the expansion for $\Phi$ (\ref{expansion}).
We write here explicitly only the two monopole terms, i.e. terms
which are independent of the angle $\chi$.

\item REG Green's --- Next we show that the piece of
$h_{zz}^{[2]}$ in (\ref{partasym2})
 involving an integral with Green's function
\[
\int \(\Phi_{,i}\Phi_{,j}\)_{REG}(r',z') G(r-r',z-z') d\,V_{d-1}.
\]
does not contain terms of the form (\ref{theterm1}). For this
purpose let us look at the leading term near $\rho=0$ in the
integral \be \int_0^{\infty} \(\Phi_{,i}\Phi_{,j}\)_{REG}(\rho')
G(\rho,\rho') \rho'^{d-2}\,d\,\rho'
 \simeq \int_{\rho}^{\infty} {1 \over \rho'^{d-3}}\, {1 \over \rho'^{d-3}}\,
  \rho'^{d-2}\,d\,\rho'
\simeq {1 \over \rho^{d-5}}~,  \label{REG-bound} \ee
 where in the first
equality we estimated the singular angle-independent part of
$\(\Phi_{,i}\Phi_{,j}\)_{REG}$ by $1/\rho'^{d-3}$ and replaced
$1/|x-x'|^{d-3}$ in the Green function by $1/\rho'^{d-3}$ with a
cutoff at $\rho$.
 Thus, we find that the leading behavior when $\rho \rightarrow 0$
is ${\cal O}\(\rho^{-(d-5)}\)$, namely there are no terms that
behave as (\ref{theterm1}). \ei

Adding up these 3 contributions for both $h_{tt}^{[2]}$ and
$h_{zz}^{[2]}$ we find exact agreement with the near zone
expressions (\ref{near1}), as seen in table \ref{table1}.
Therefore the matching constants indeed vanish.

\subsection{Results}
\label{results}

We did not find an analytic solution for the integral
(\ref{partasym2}), and hence we do not have a full analytic form
for the metric at this order. However, we can analytically
evaluate the asymptotic form of this metric, thus determining the
PN corrections to the thermodynamic quantities, $M,\, \tau$. The
results agree with Harmark's results \cite{H4} which were obtained
using a different, single patch, method.

Our goal is to find $c_{tt}^{[2]}$ and $c_{zz}^{[2]}$, the next to
leading corrections to $g_{tt}$ and $g_{zz}$ for $r \gg z$. From
the asymptotic form (\ref{Phiasym}) we see that the terms in the
expansion of $\Phi^{2}$ for $r \gg z$ do not contribute to the
asymptotic constants. Moreover, we found in the last subsection
that the matching constants vanish. Thus only the source terms
(the RHS in (\ref{post})) contribute and we have \bea
c_{tt}^{[2]}&=&0, \\
c_{zz}^{[2]}&=&\frac{d-2}{2\, (d-3)\,(d-4)\,
\Omega_{d-3}}\,\mbox{Pf}\(\int\(\Phi_{,z}\)^{2} d\,V_{d-1}\).
\label{czz2} \eea $\mbox{Pf}$ stands for the finite part of the
divergent integral according to Hadamard's partie finie
regularization.

On dimensional grounds \be \mbox{Pf}\(\int\(\Phi_{,z}\)^{2}
d\,V_{d-1}\)=const\, {\rho_0^{~2(d-3)} \over L^{d-3}}~, \ee where
$const=const(d)$. We evaluated these constants by separating the
integral into SI and REG pieces. The partie finie of the SI part
vanishes and we find (the detailed calculation of the integral and
its regularization appears in Appendix \ref{integral}) \be
\mbox{Pf}\(\int\(\Phi_{,z}\)^{2}
d\,V_{d-1}\)=-\frac{\rho_{0}^{2\,(d-3)}}{L^{d-3}}\Omega_{d-2}(d-4)(d-3)\zeta(d-3)~.
\ee
 Substituting back into (\ref{czz2}) gives us the final expressions \bea
c_{tt}^{[2]}&=&0, \\
c_{zz}^{[2]}&=&-\frac{\rho_{0}^{2\,(d-3)}}{L^{d-3}}\,
\frac{\zeta(d-3)\,(d-2)\, \Omega_{d-2}}{2\, \Omega_{d-3}}. \eea
Using (\ref{mass},\ref{tens}) and adding it to the results in the
Newtonian approximation (\ref{firstmass}) we obtain \emph{the
``measurable'' quantities} as

\be \fbox{$ \room ~~\begin{array}{ccc}
 \room M&=&\frac{(d-2)\Omega_{d-2}}{16 \pi\,G_{d}}\,\rho_{0}^{d-3}
 \(1+\frac{\zeta(d-3)}{2}\,\frac{\rho_{0}^{d-3}}{L^{d-3}}\)
 +\mathcal{O}\(\rho_{0}^{3(d-3)}\)~~~, \label{secondmass}\\
 \room \tau\, L &=&\frac{(d-3)\,(d-2)\,\Omega_{d-2}\,\zeta(d-3)}{32\pi\,
G_{d}}\,\frac{\rho_{0}^{2\,(d-3)}}{L^{d-3}}+\mathcal{O}\(\rho_{0}^{3(d-3)}\)~~.
\end{array} ~~$}
\label{firsttension} \ee

\subsection{Newtonian derivation of tension to leading order}
\label{Newtonian-tension}

We shall now show that the leading order tension
(\ref{firsttension}) can be understood from Newtonian gravity.
From the first law of black hole thermodynamics (\ref{first-law})
 we can express the tension as \be
\tau=\(\frac{d\,M}{d\,L}\)_{S} ~, \ee
 namely, the change of the
mass in reaction to a change in the period of the compact
dimension when we keep the entropy constant. The demand for
constant entropy of a small black hole just means that $\rho_0$
(as well as the rest-mass $M_0$) are fixed. The black hole mass in
the Newtonian limit is \be
 M = M_0+U ~,\ee
where the mass is expressed as a sum of $M_0$, the intrinsic mass
 of the black hole given by the leading order of (\ref{secondmass})
\be
 M_0 = \frac{(d-2)\Omega_{d-2}}{16 \pi\,G_{d}}\,\rho_{0}^{d-3}
 \label{rest-mass} ~,\ee
 $U \ll M_0$ is
 the gravitational potential energy, written as a sum over
all mirror images with the standard prefactor $1/2$ to avoid
over-counting \be
 U =\frac{1}{2}\!\sum_{n\,\in\, \mathbb{Z}\backslash
\{0\}}\!\!\!\!U_{n} = \sum_{n\,\in\, \mathbb{N}}\, U_{n} ~.\ee

The (negative) Newtonian gravitational potential energy is given
by \be
 U=-\half\, h_{tt}\, M_0 ~. \label{U-htt} \ee
 We may motivate this expression from two perspectives: geodesic
motion or the red-shift of energy.

From the geodesic motion point of view $U$ is obtained by
integration over the Newtonian force that is needed to bring the
black hole from infinity. The Newtonian force is determined from
the Geodesic equation \be
\ddot{x}^{\alpha}+\Gamma_{\mu\nu}^{\alpha}\,\dot{x}^{\mu}\dot{x}^{\nu}=0,
\ee that in the Newtonian limit (weak field limit) gives \be
\frac{d^{2}\,x^{i}}{d\,t^{2}}=\frac{1}{2}\,h^{[1]}_{tt,i}~, \ee
where $h^{[1]}_{tt}$ is given in (\ref{Newton1}). Comparing with
Newton's second law we write the Newtonian gravitational field as
\be \vec{E}=\frac{1}{2}\,\nabla h_{tt}, \ee thereby motivating
(\ref{U-htt}).

Alternatively, the energy red-shift is given by \bea
 U &=& M-M_0 \approx \del_t-\del_\tau \approx \(1- \sqrt{-g^{tt}}\) \, \del_t
 \non
  &\approx&  \(1- \sqrt{-g^{tt}}\)\,M_0=
  \(1-\sqrt{1+h_{tt}}\)\,M_{0}
  \simeq -\half\, h_{tt}\, M_0
  ~.\eea

Now we may substitute
\[U_{n}=-\frac{1}{2}\,\frac{\rho_{0}^{d-3}\,M_0}{{(n\,L)}^{d-3}},\]
and find \be \tau\, L = L\, \frac{d\,U}{d\,L}=-L\, {d \over dL}
\sum_{n\,\in\,
\mathbb{N}}\frac{\rho_{0}^{d-3}\,M_0}{2\,n^{d-3}\,L^{d-3}}=
\frac{(d-3)\,\zeta(d-3)}{2}\, \frac{\rho_{0}^{d-3}}{L^{d-3}} ~
M_0. \ee Finally, substituting the expression for the rest mass
(\ref{rest-mass}) we recover the tension formula
(\ref{firsttension}).

\section{Implications for the phase diagram}
\label{implications}

In this section we shall translate the PN thermodynamic constants
which we obtained into the phase diagram. The tension formula will
dictate the slope of the small black hole branch. By extrapolation
we shall attempt to learn about the phase transition region.
Following the prediction for a critical dimension in this system
\cite{TopChange}, we shall explore the dimensional dependence of
the extrapolation and indeed we shall find an indication for
Sorkin's critical dimension \cite{SorkinD*}.\footnote{This
extrapolation could have been done ever since the results were
first found in \cite{H4}.}

For the phase diagram it is convenient to define a dimensionless
mass to serve as a control parameter
\[\mu:= \frac{G_{d}\, M}{L^{d-3}}.\]
 From the expression for the tension (\ref{firsttension})
 we obtain an estimated $\tmu$ as a function of the dimensionless tension $n$
for small black holes \be
\tmu_{BH}=\frac{(d-2)\Omega_{d-2}}{8\,\pi\,(d-3)\,\zeta(d-3)}\,n,
\label{themu} \ee where the dimensionless tension $n$ is defined
by
\[n := \frac{\tau\,L}{M}.
\]
This quantity serves us as a good dimensionless order parameter
for the phase diagram~\cite{KPS1,HO1}, and on the uniform
black-string branch it attains the constant value \be
n_{st}=\frac{1}{d-3}. \ee

\begin{figure}[htb]
\epsfxsize=0.7\textwidth
 \centerline{\epsffile{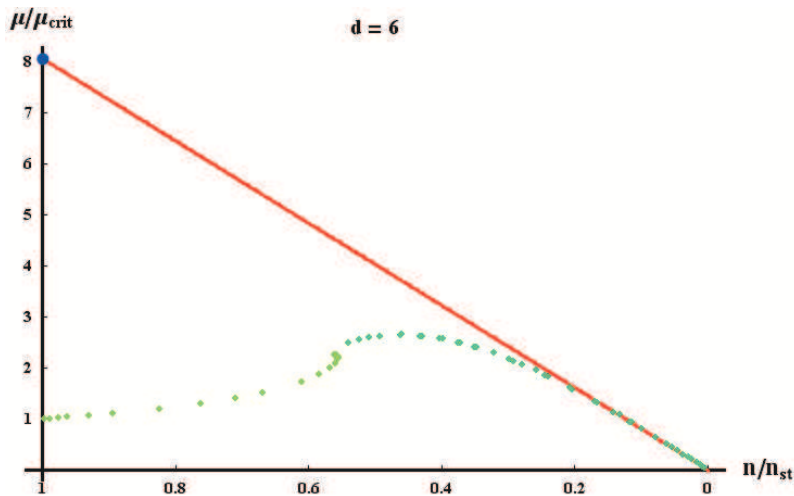}}
\caption{\small The analytic linear approximation line (red)
compared to the numerical data points (green) of
\cite{KudohWiseman2} for the complete phase diagram in $d=6$
consisting of two branches: the black hole and the non-uniform
black string. The vertical axis is proportional to the mass, and
the horizontal axis to the tension (see text for exact
definitions). For small black holes (lower right) we see excellent
agreement. By extrapolation, the thick (blue) point defines
$\mu_X$, the intersection point of the extrapolated linear
approximation with the line of uniform black strings, $n=n_{st}$.}
 \label{inter6}
\end{figure}

In figure \ref{inter6} we give the mass $\mu$ against the tension
$n$ for $d=6$ for both the analytic linear approximation
(\ref{themu}) and the numerical results for the whole phase
diagram (the black hole and the non-uniform black string) taken
from \cite{KudohWiseman2} (where the results of
\cite{KudohWiseman1,Wiseman1} are incorporated\footnote{We are
grateful to T. Wiseman and H. Kudoh for sending us their data.}) .
For the small black hole branch (lower right) we see excellent
agreement.

\begin{figure}[htb]
\epsfxsize=0.63\textwidth
 \centerline{\epsffile{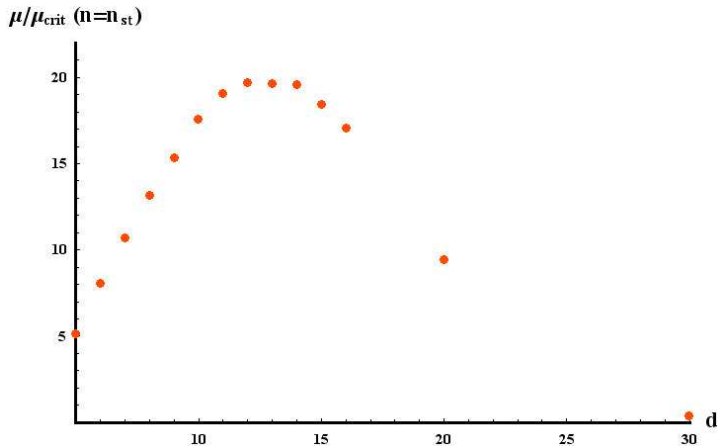}}
\caption{\small The ratio $\mu_X/\mu_{crit}$ of extrapolated vs.
actual Gregory-Laflamme points for various dimensions. For large
$d$ the ratio tends to zero strongly. We interpret that as an
indication for a higher (second) order phase transition, and thus
as an indication for Sorkin's critical dimension $D^*=``13.5"$.
Interestingly, our graph shows a maximum around $12 \le d \le 14$
which coincides with $D^*$.}
 \label{interd}
\end{figure}

Next we would like to extrapolate and define the extrapolated
intersection point of the linear approximation with the black hole
branch \be
 \mu_X := \tmu_{BH}(n_{st})=\frac{(d-2)\Omega_{d-2}}{8\,\pi\,(d-3)^2\,\zeta(d-3)}
 \label{def-muX} ~,\ee
 (see also figure \ref{inter6}). Now we can compare $\mu_X$ with the Gregory-Laflamme (GL)
critical mass $\mu_{crit}$ as a function of the dimension. In
figure \ref{interd} we plot $\mu_X/\mu_{crit}$ for various $d$.
The critical mass was computed from $k_{GL}$, the critical
wave-number through \be
\mu_{crit}=\frac{(d-3)\,\Omega_{d-3}}{16\,\pi}\,\(\frac{k_{GL}}{2\,\pi}\)^{d-4}~,
\ee and values of $k_{GL}$ for various dimensions were obtained in
\cite{SorkinD*}  and can be found in \cite{KolS}.

For large $d$, $k_{GL}$ behaves as $k_{GL}\sim \sqrt{d}$ and the
ratio $\mu_X/\mu_{crit}$ tends to zero strongly. A similar
observation appears in \cite{HO-review}. We interpret that as an
indication for a second order phase transition, where the black
hole phase does not reach the Gregory-Laflamme region, but rather
turns first into a stable non-uniform string phase, and only the
latter joins into the GL point. Since \cite{Gubser} it is known
that at $d=5$ the transition is first order. Thus we find an
indication for a critical dimension, which separates first and
second order behavior and was indeed discovered by Sorkin to lie
at $D^*=``13.5"$ \cite{SorkinD*}. Interestingly, our graph shows a
maximum in the range $12 \le d \le 14$ which is the location of
the critical dimension $D^*$.

\vspace{0.5cm} \noindent {\bf Acknowledgements}

We would like to thank H. Kudoh, E. Sorkin and T. Wiseman for
sharing their numerical results and for discussions. We would like
to thank J. Avron, S. Elitzur and especially A. Ori for
discussions.

This work is supported in part by The Israel Science Foundation
(grant no 228/02) and by the Binational Science Foundation
BSF-2002160.

\appendix
\section{The Calculation of $\mbox{Pf}\(\int\(\Phi_{,z}\)^{2}
d\,V_{d-1}\)$} \label{integral}

We start by replacing $\Phi$ by a sum over images
 \bea &&\mbox{Pf}\(\int\(\Phi_{,z}\)^{2}
d\,V_{d-1}\)= \label{A1} \\
&&=\mbox{Pf}\((d-3)^{2}\rho_{0}^{2\,(d-3)}\!\!\!\!\sum_{(m,n) \in
\,\IZ^{2}} \!\int_{- \frac{L}{2} }^{\frac{L}{2}}\!\! d\,z
\int_{0}^{\infty}\!\! r^{d-3}\,d\,r
\frac{\(z+m\,L\)\,\(z+n\,L\)}{\(r^{2}+(z+m\,L)^{2}\)^{\frac{d-1}{2}}\,\(r^{2}+(z+n\,L)^{2}\)^{\frac{d-1}{2}}}\).\nonumber
\eea
 The summation here goes over all the pairs of mirror images
of the black hole, and we transform it as follows \be
 \sum_{(m,n) \in \,\IZ^{2}} \!\int_{-\frac{L}{2}}^{\frac{L}{2}} \!\!
 d\,z = \sum_{k=m-n \in \,\IZ} \int_{-\infty}^{+\infty} dz
 =\left[~ |_{k=0}+2 \sum_{k=1}^{\infty} ~\right] \int_{-\infty}^{+\infty} dz ~,\ee
where according to (\ref{sing2},\ref{reg2}) \bea
 |_{k=0}&=& \(\Phi_{,z}^{~~2}\)_{SI},\\
 2 \sum_{k=1}^{\infty} &=& \(\Phi_{,z}^{~~2}\)_{REG}.
 \eea

This transformation, replacing the integration over
$-\frac{L}{2}<z<\frac{L}{2}$ by integration over the covering
space $-\infty<z<\infty$, is justified since the integral is
invariant under translations in the direction of the periodic
coordinate $z$, and thus we can replace summation over images with
fixed $k \equiv m-n$ by a summation of integrals over stripes
$(m-1/2)L \le z \le (m+1/2)L$ which altogether exhaust the whole
range of $z$ in the covering space: for fixed $k$ we have $\sum_m
\int_{-L/2}^{+L/2} dz\, I_{m,n} = \sum_m
\int_{(m-1/2)L}^{(m+1/2)L} dz\, I_{k,0}$, where $I_{m,n}$ is the
integrand in (\ref{A1}). Then we wrote the sum over $k$ as a sum
over natural numbers $k\in \,\IN$ (and multiplied by two) when
$k\,L$ is the distance of the black hole from its mirror image,
and we separated the singular term which corresponds to $k=0$.

Now, we have to regularize just the singular term that comes from
the self-interaction  \be
 \mbox{Pf}\(\int \Phi_{,z}^{~~2}\, dV_{d-1}\) =
 \mbox{Pf}\(\int\(\Phi_{,z}^{~~2}\)_{SI}\, dV_{d-1} \)
            +\int\(\Phi_{,z}^{~~2}\)_{REG}\, dV_{d-1}. \ee
 This is done according to the
idea of Hadamard's partie finie regularization introduced in
section \ref{divergences}. Transforming the singular term to polar
coordinates $(\rho,\chi)$ and imposing a cut-off at
$\rho=\varepsilon$ we see that it is proportional to
\[\int_{\varepsilon}^{\infty}\frac{d\rho}{\rho^{d-2}} \propto\frac{1}{\varepsilon^{d-3}}~,\]
namely \be \mbox{Pf}\(\int\(\Phi_{,z}^{~~2}\)_{SI}\, dV_{d-1} \)=
0 ~.\ee

We turn to the regular part \bea
 && \int\(\Phi_{,z}^{~~2}\)_{REG}\, dV_{d-1} =\non
 &=& 2\,\rho_{0}^{2\,(d-3)}\,(d-3)^{2}
 \sum_{k\in \,\IN}\!\int_{-\infty }^{\infty}\!\!
  d\,z \int_{0}^{\infty}\!\! r^{d-3}\,d\,r\,
\frac{z}{\(r^{2}+z^{2}\)^{\frac{d-1}{2}}}\cdot\frac{\(z+k\,L\)}
{\(r^{2}+(z+k\,L)^{2}\)^{\frac{d-1}{2}}}. \eea
 Integration by parts
 gives \be
2\,\rho_{0}^{2\,(d-3)}\,(d-3)\sum_{k\in \,\IN}\!\int_{- \infty
}^{\infty}\!\! d\,z \int_{0}^{\infty}\!\! r^{d-3}\,d\,r
\,\frac{1}{\(r^{2}+z^{2}\)^{\frac{d-3}{2}}}\cdot
\frac{\partial}{\partial
z}\(\frac{\(z+k\,L\)}{\(r^{2}+(z+k\,L)^{2}\)^{\frac{d-1}{2}}}\).\ee
This integral can be rewritten in the form \be
-2\,\rho_{0}^{2\,(d-3)}\sum_{k\in
\,\IN}\left.\frac{\partial^{2}}{\partial
b^{2}}\right|_{b=k\,L}\!\int_{- \infty }^{\infty}\!\! d\,z
\int_{0}^{\infty}\!\! r^{d-3}\,d\,r \,
\frac{1}{\(r^{2}+z^{2}\)^\frac{d-3}{2}}\cdot
\frac{1}{\(r^{2}+(z+b)^{2}\)^\frac{d-3}{2}}=,\ee transforming to
polar coordinates and introducing a dimensionless variable
$u:=\frac{\rho}{b}$ leads us to \be
=-2\,\rho_{0}^{2\,(d-3)}\Omega_{d-3}\sum_{k\in
\,\IN}\left.\frac{\partial^{2}}{\partial b^{2}}\right|_{b=k\,L}
\frac{1}{b^{d-5}}\int_{0}^{\infty}u \,d\,u
\int_{0}^{\pi}d\,\chi\frac{\sin^{d-3}(\chi)}{\(u^{2}+2\,u
\cos(\chi)+1\)^{\frac{d-3}{2}}}.\ee

The integral over the angle $\chi$ gives \bea
\int_{0}^{\pi}d\,\chi\frac{\sin^{d-3}(\chi)}{\(u^{2}+2\,u
\cos(\chi)+1\)^{\frac{d-3}{2}}}=\left\{\begin{array}{lr}
\frac{\Omega_{d-2}}{\Omega_{d-3}} \qquad \qquad 0<u\leq1\\
\frac{\Omega_{d-2}}{\Omega_{d-3}}\cdot \frac{1}{u^{d-3}} \quad \;
1<u<\infty
\end{array} \right..\eea
Finally, performing the $b$-derivatives and integrating over $u$
gives us the final result \be \mbox{Pf}\(\int\(\Phi_{,z}\)^{2}
d\,V_{d-1}\)=-\frac{\rho_{0}^{2\,(d-3)}}{L^{d-3}}\Omega_{d-2}(d-4)(d-3)\zeta(d-3).
\ee

\section{The next to leading correction to the ``Archimedes'' effect}
\label{Archimedes}

In \cite{previous} we computed the leading order for the
inter-polar distance defined to be the proper distance between the
``poles'' of the black hole measured around the compact dimension
\be \Lp =2\, \int_{z_H}^{L/2} dz\, \sqrt{g_{zz}}, \ee where $z_H$
denotes the location of the horizon. It is convenient to define
the dimensionless quantities \bea
 y: &=& 1-{\Lp \over L} ~, \label{def-y} \\
 \eta &:=& {\rho_H \over L} ~, \label{def-eta}\eea
 where $\rho_H$ is the location of the horizon in isotropic
 coordinates, related to the \Schw ~radius, $\rho_0$ through \be
 \rho_0^{~d-3} = 4\, \rho_H^{~d-3}~. \label{rhoH-rho0} \ee
 In \cite{previous} we found \be
 y = 2\, I_d\, \eta + o(\eta)~, \label{leading-Arch} \ee
 where the constants $I_d$ are defined by \be
 I_d :=  1- \int_0^1 \( \(1+w^{d-3}\)^{{2 \over
d-3}} -1 \)\,
 {dw \over w^2} = 4^{1/k}\, \sqrt{\pi}\,
 {\Gamma\({k-1 \over k}\)
 \over \Gamma\({1 \over 2} - {1 \over k}\)}, ~~ k=d-3~,
 \label{def-Id} \ee
 and are a monotonic function of $d$: $I_5=0,\, I_6=0.6845,\,
I_\infty=1$.

Here we shall compute further corrections for $y$, getting up to
order $d-2$ in $\eta$. Let us start by identifying the necessary
orders at each patch. In order to compute $y$ at order $\eta^s$ it
is necessary to know the asymptotic metric up to order $s$, but in
the near zone it is sufficient to know the metric up to order
$s-1$, due to the division by $L$ in the definition (\ref{def-y}).
Thus for the leading result (\ref{leading-Arch}) it is sufficient
to use the asymptotic zone metric up to order 1, which is the same
as order 0, namely the flat metric, and in the near zone up to
order 0, namely \Schw . Here we will use our information about the
metric in both zones up to order $d-3$, which includes the
Newtonian approximation in the asymptotic zone, and the first
(monopole) correction in the near zone. Since the next order in
the asymptotic zone always vanishes, our result will hold up to
$\eta^{d-2}$. \footnote{In principle, we have enough information
to get up to order $2(d-3)-1$, since there are no corrections in
the asymptotic zone up to that order and in the near zone there
are known corrections from matching with the Newtonian potential
(see subsection \ref{near-zone-rev}). However, we did not perform
this computation since in 5d, the only order where we do not know
yet the leading behavior since $I_5=0$, order $2(d-3)-1=3$
coincides with order $d-2$ which we compute here, and thus such a
higher order computation would add information only at higher
dimensions where it is less needed.}

The way to compute $L_{poles}$ is to pick some mid-point $Z$,
divide the integration between the two zones\be
 \half\, \Lp =\int_{z_H}^{Z}dz\, \sqrt{g_{zz}^{\mbox{(near)}}}
 + \int_{Z}^{L/2}dz\, \sqrt{g_{zz}^{\mbox{(asymp)}}}, \label{sum-of-patches} \ee
 and confirm that the result is independent of the choice of
 mid-point.

In the asymptotic zone we have (\ref{Newton2},\ref{thephi})
  \be
 \room g_{zz}^{\mbox{(asymp)}}=1+\frac{\Phi}{d-3} +
 \mathcal{O}\left(\rho_{H}^{2\,(d-3)}\right) ~.\ee
 Thus the contribution to $\Lp/(2L)$ (\ref{sum-of-patches})
is \bea
 && \({\Lp \over 2L}\)_{\mbox{asymp}}:={1 \over L}\, \int_{Z}^{L/2}dz\, \sqrt{g_{zz}^{\mbox{(asymp)}}}=  \nonumber\\
 &=& \frac{1}{L}\int_{Z}^{\frac{L}{2}}dz\,\left(1+\sum_{n=-\infty}^{\infty}
 \frac{\rho_{0}^{d-3}}{2(d-3)\,\mid\! z+n\,L\!\mid^{d-3}
}\right)+\mathcal{O}\left(\rho_{0}^{2\,(d-3)}\right) =  \non
 &=& \half - {Z \over L} + {\rho_0^{d-3} \over 2(d-3)(d-4)\, L^{d-3}}\,
 \left[ {1 \over \tz^{d-4}}+ \sum_{n=1}^{\infty}{1 \over (n+\tz)^{d-4}}
 - \sum_{n=1}^{\infty}{1 \over (n-\tz)^{d-4}}\right]^{Z/L}_{1/2}
 + \non
 && + \mathcal{O}\left(\rho_{0}^{2\,(d-3)},\right)
 \eea
where in the last equality we defined $\tz:=z/L$. For $\tz=1/2$
the expression in brackets in the last line vanishes. At the other
boundary, $\tz=Z/L$, we should expand up to order $\tz$ (since we
need terms up to order $L^{-(d-2)}$ in the overlap region. Note
that here is quite easy to extend the result to order $2(d-3)-1$,
which we discussed above, simply by expanding further). Altogether
we find  \be
 \({\Lp \over 2L}\)_{\mbox{asymp}} = \half - {Z \over L} +
 {\rho_0^{d-3} \over 2(d-3)(d-4)\, L^{d-3}}\,
 \left[ \({L \over Z}\)^{d-4} \!\!\! - 2(d-4)\, \zeta(d-3)\, {Z \over L}
 \right]. \label{Arch-asymp} \ee

We now turn to the near zone. Here at order $(d-3)$ we consider
only the monopole correction, which leaves us with the \Schw
~metric, with adjusted parameters. For matching purposes it is
convenient to write the \Schw ~metric in isotropic coordinates \be
 ds^{2} = - \( {1 - \psi \over 1 + \psi} \)^2\, dt^2 + (1 +
\psi)^{4 \over d-3}\, \( d\rho_{c}^2 + \rho_{c}^2\, d
\Omega^2_{d-2} \), \ee where \be
 \psi = \( \rho_H \over \rho_{c} \)^{d-3} ~.\ee

To incorporate the correction we could have used (\ref{near1}) in
a straightforward manner, but here we take an alternative route.
 In order to match we are free to adjust $\rho_H$, rescale $t$ and
re-parametrize $\rho_c$. Since we wish to retain the isotropic
form we consider only a rescaling of $\rho_c$. Matching of
$g_{zz}$ is achieved by the transformation \bea
 \rho_c &\to& (1+ \delta)\, \rho_c~, \non
 \delta &:=& {1 \over 2(d-3)}\, \Phi_0 =
 {\zeta(d-3) \over d-3}\, \({\rho_0 \over L}\)^{d-3}~, \label{match-adj1} \eea
 where $\Phi_0$ is the Newtonian potential at the origin due to
 the images. Matching of $g_{tt}$ requires also \be
 t \to (1+(d-3)\, \delta)\, t ~.\ee
 One is free to change $\rho_H$ as well -- this is a change of
scheme, or reparametrization of the branch of solutions. Here it
is convenient to choose $\rho_H$ to remain unchanged \be
 \delta\rho_H=0 \label{scheme} ~,\ee
(compare this scheme with the scheme of (\ref{near1}) which
guarantees zero matching at PN order and in the current context
would be expressed as
 $\rho_H \to \rho_H\, (1+\delta)$.)

We may now calculate the contribution of the near zone to
$\Lp/(2L)$ (\ref{sum-of-patches}) \bea
 && \({\Lp \over 2L}\)_{\mbox{near}} :=
 {1 \over L}\, \int_{z_H}^{Z} dz\, \sqrt{g_{zz}^{\mbox{(near)}}}=  \non
 &=& {1 \over L}\, \int_{z_H}^{Z} dz\, \(1+\psi\)^{2/(d-3)} = \non
 &=& {\rho_H \over L}\, \int_{1/(1+\delta)}^{Z/\rho_H}
 \(1+\({\rho_H \over z\, (1+\delta)}\)^{d-3} \)^{2/(d-3)}
 (1+\delta){dz \over \rho_H} = \non
 &=& {\rho_H \over L}\, \int_1^{Z(1+\delta)/\rho_H}
 \(1+\({\rho_H \over z\, }\)^{d-3} \)^{2/(d-3)}
 {dz \over \rho_H} = \non
 &=& {\rho_H \over L}\, \left[ -I_d + { Z\,(1+\delta) \over
 \rho_H} -
 \int_0^{\rho_H/(Z(1+\delta))} \( (1+t^{d-3})^{2/(d-3)}-1 \) \, {dt \over t^2}
 \right],
 \eea
where in passing from the second to the third lines we used the
adjustments (\ref{match-adj1}), including the adjustment in the
location of the horizon induced by the rescaling of $\rho_c$; in
the next line we performed a change of variables $(1+ \delta)\, z
\to z$ and in passing to the fifth we changed to $t:=\rho_H/z$.
Finally expanding the integral up to order $\rho_H^{d-3}$ we find
the total contribution from the near zone to be \be
 \({\Lp \over 2L}\)_{\mbox{near}} =
  {\rho_H \over L}\, \left[ -I_d +{ Z\,(1+\delta) \over
 \rho_H}-{2 \over (d-3)(d-4)} \({\rho_H \over
 Z\,(1+\delta)}\)^{d-4}
 \right]~. \label{Arch-near} \ee

Adding up (\ref{Arch-asymp}) and (\ref{Arch-near}) we find that
indeed all the $Z$ dependence disappears to the prescribed order
once we recall the definition of $\delta$ (\ref{match-adj1}) and
the relation between $\rho_H,\rho_0$ (\ref{rhoH-rho0}), and we are
left with the final answer \be  {\Lp \over 2L} = \half -\eta\, I_d
+o\( \({\rho_0 \over L}\)^{d-2}\)~,\ee or equivalently
\be \fbox{\room$~~~
 y=2\, I_d\, \eta +o\( \({\rho_0 \over L}\)^{d-2}\).  ~~$} \label{Arch-high}\ee

Summarizing, our result is expressed in terms of $y,\, \eta,\,
I_d$ whose definitions  are given in
(\ref{def-y},\ref{def-eta},\ref{def-Id}). It turns out that using
the scheme choice (\ref{scheme}) our higher order result
(\ref{Arch-high}) is precisely  our low order result
(\ref{leading-Arch}). In particular we find that in 5d $y$
vanishes up to order $d-2=3$ in $\eta$ (this is scheme
independent) which is consistent with 5d numerical simulations
\cite{KPS2} where $y$ appears to be ${\cal O}(\eta^4)$ (however,
the result in \cite{KSSW1} is different).

\end{document}